\tolerance=10000
\input phyzzx

\REF\CW{C.M. Hull and N. P. Warner, Class. Quant. Grav. {\bf 5} (1988)
1517.}
\REF\gib{ G.W. Gibbons, \lq Aspects of Supergravity Theories', Published
in {\it GIFT Seminar on  Supersymmetry, Supergravity and Related Topics},
edited by F. del Aguila, J. A. de Azcarraga and L. E. Ibanez, World
Scientific (1984). }
\REF\mn{J. Maldacena and C. Nunez, {\it  Supergravity description of
field theories on curved manifolds and a  no go theorem} Int.J.Mod.Phys.
{\bf A16} (2001) 822; hep-th/0007018.}
\REF\wall{C.M. Hull, {\it Domain Wall and 
de Sitter Solutions of Gauged Supergravity}, hep-th/0110048.}
\REF\CJ{E. Cremmer and B. Julia, Phys. Lett. {\bf 80B} (1978) 48; Nucl.
Phys. {\bf B159} (1979) 141.}
\REF\dwn{ B. de Wit and H. Nicolai,  {\it Phys. Lett.} {\bf 108 B} (1982)
285; B. de Wit and H. Nicolai,  {\it Nucl. Phys.} {\bf B208} (1982) 323.}
\REF\nct{C.M. Hull, {\it Phys. Rev.} {\bf D30} (1984) 760; C.M. Hull,
{\it Phys. Lett.} {\bf 142B} (1984) 39; C.M. Hull, {\it Phys. Lett.} {\bf
148B} (1984) 297; C.M. Hull, {\it Physica} {\bf 15D} (1985) 230; Nucl.
Phys. {\bf B253} (1985) 650.}
\REF\nctt{ 
  C.M. Hull, {\it Class. Quant. Grav.} {\bf
2} (1985) 343.}
\REF\DS{C.M. Hull, {\it 
De Sitter space in supergravity and M theory},
hep-th/0109213.}
\REF\HW{C.M. Hull and N. P. Warner,  { Nucl. Phys.} {\bf B253}
(1985) 650, 675.}
\REF\gunwar{ M. G\"unaydin, L.J. Romans and N.P. Warner, {\it Gauged $N=8$
Supergravity in Five Dimensions}, Phys. Lett. {\bf 154B}, n. 4 (1985)
268; {\it Compact and Non--Compact Gauged Supergravity Theories in Five
Dimensions}, Nucl. Phys. {\bf B272} (1986) 598.}
\REF\frefo{ F. Cordaro, P. Fr\'e, L. Gualtieri, P. Termonia and M.
Trigiante, {\it $N=8$ gaugings revisited: an exhaustive classification},
Nucl. Phys. {\bf B532} (1998) 245.}
\REF\PPV{ M. Pernici, K. Pilch and P. van Nieuwenhuizen, {\it Gauged
$N=8~D=5$ Supergravity} Nucl. Phys. {\bf B259} (1985) 460.}
\REF\popec{ M.~Cvetic, H.~Lu, C.~N.~Pope, A.~Sadrzadeh and T.~A.~Tran,
{\it `S(3) and S(4) reductions of type IIA supergravity,} Nucl.\ Phys.\ B
{\bf 590}, 233 (2000) hep-th/0005137.}
\REF\vanW{P. van Nieuwenhuizen and N. P. Warner, Commun. Math. 
Phys. {\bf 99} (1985) 141.}
\REF\PPVa{ M. Pernici, K. Pilch and P. van Nieuwenhuizen, Nucl. Phys. {\bf
B249} (1985) 381.}
\REF\ans{ B. de Wit, H. Nicolai and N. P. Warner,    { Nucl. Phys.}
{\bf B255} (1985) 29.}
\REF\consis{ B. de Wit and H. Nicolai,    { Nucl. Phys.} {\bf B281}
(1987) 211.}
\REF\randsum{L. Randall and R. Sundrum, Phys. Rev. Lett. {\bf 83} (1999) 3370;
4690.}
\REF\gel{M. Gell-Mann and B. Zwiebach,  Phys. Lett. {\bf 141B} (1984) 333, {\bf
147B} (1984) 111; C. Wetterich, Nucl. Phys. {\bf 242} (1984) 473.}
\REF\NicWet{
H.~Nicolai and C.~Wetterich,
{\it On The Spectrum Of Kaluza-Klein Theories With Noncompact Internal
Spaces}, Phys.\ Lett.\ B {\bf 150}, 347 (1985).}
\REF\LPS{ H. L\"{u}, C.N. Pope, E. Sezgin and K.S. Stelle,
{\sl Dilatonic p-brane solutions}, Phys. Lett. {\bf B371} (1996) 46-50,
hep-th/9511203.}
\REF\stains{ H. L\"{u}, C.N. Pope, E. Sezgin and K.S. Stelle,
Nucl.  Phys. Lett. {\bf B456} (1995) 669.}
\REF\CowdallTW{
P.~M.~Cowdall, H.~Lu, C.~N.~Pope, K.~S.~Stelle and P.~K.~Townsend,
{\it Domain walls in massive supergravities},
Nucl. Phys. B {\bf 486}, 49 (1997)
[hep-th/9608173].}
\REF\LuRH{
H.~Lu, C.~N.~Pope and P.~K.~Townsend,
{\it Domain walls from anti-de Sitter spacetime},
Phys.\ Lett.\ B {\bf 391}, 39 (1997)
[hep-th/9607164].}
\REF\CveticXX{
M.~Cvetic, S.~S.~Gubser, H.~Lu and C.~N.~Pope,
{\it Symmetric potentials of gauged supergravities in diverse dimensions 
and  Coulomb branch of gauge theories},
Phys.\ Rev.\ D {\bf 62}, 086003 (2000)
[hep-th/9909121].}
\REF\Behrn{
K.~Behrndt, E.~Bergshoeff, R.~Halbersma and J.~P.~van der Schaar,
{\it On domain-wall/QFT dualities in various dimensions},
Class.\ Quant.\ Grav.\  {\bf 16}, 3517 (1999)
[hep-th/9907006].}
\REF\Ahn{
C.~Ahn and K.~Woo,
{\it Domain wall and membrane flow from other gauged d = 4, n = 8 
supergravity. I}, [hep-th/0109010].}
\REF\Gibbwilt{
G. W. ~Gibbons and D. L.  ~Wiltshire, {\it Spacetime as
a membrane in higher dimensions}, { Nucl. Phys. } {\bf B
287}(1987) 717 [hep-th/010109093].}
\REF\Gibbons{
G.W. ~Gibbons, ``Quantized Fluxtubes in Einstein-Maxwell
Theory and Noncompact Internal spaces'', in { \it Fields and
Geometry, Proceedings of 22nd Karpacz Winter School } (1986)
517-613, ed A Jadczyk,  World Scientific. }
\REF\Gibb{
G. W. ~Gibbons,  ``The Dimensionality of
Spacetime '', in {\it String Theory, Quantum Cosmology and
Gravity, Integrable and Conformal Invariant Theories},
Proceedings of the Paris-Meudon Colloquium (1986) 16-25, eds H J
de Vega and N Sanchez,  World Scientfic. }
\font\mybb=msbm10 at 12pt
\def\bbbb#1{\hbox{\mybb#1}}

\def\R {\bbbb{R}}
 %

\def \aa {\alpha}
\def \bb {\beta}

\def \dd {\delta}

\def \ll {\lambda}
\def \mm {\mu}
\def \nn {\nu}

\def \ss {\sigma}

\def \th {\theta}

\def \lll {\Lambda}

\def \www{\Omega}

\def \ti {\tilde}

\def \2 {{1 \over 2}}
\def \3 {{1 \over 3}}
\def \4 {{1 \over 4}}
\def \5 {{1 \over 5}}
\def \6 {{1 \over 6}}
\def \7 {{1 \over 7}}
\def \8 {{1 \over 8}}
\def \9 {{1 \over 9}}
\def \0 { \infty}

\def\++ {{(+)}}
\def \- {{(-)}}
\def\+-{{(\pm)}}

\def \qq {\qquad}


 \def\unit{\hbox to 3.3pt{\hskip1.3pt \vrule height 7pt width .4pt
\hskip.7pt
\vrule height 7.85pt width .4pt \kern-2.4pt
\hrulefill \kern-3pt
\raise 4pt\hbox{\char'40}}}
\def\II{{\unit}}

\def\H {{\cal{H}}}

\def\nup#1({Nucl.\ Phys.\  {\bf B#1}\ (}



\Pubnum{ \vbox{  \hbox {QMUL-PH-01-13}   \hbox {DAMTP-2001-100} 
\hbox{hep-th/0111072}  
}}
\pubtype{}
\date{November 2001}

\titlepage

\title {\bf  De Sitter Space from Warped 
Supergravity Solutions}

\author {G.W. Gibbons}
\address{DAMTP, Centre for Mathematical Sciences, Wilberforce Road, \break
Cambridge CB3
OWA, UK.}

\andauthor{C.M. Hull}
\address{Physics Department,
\break Queen Mary, University of London,
\break Mile End Road, London E1 4NS, U.K.}

\vskip 0.5cm

\abstract { The solutions of 10 and 11 dimensional supergravity that are
warped products of de Sitter space with a non-compact `internal' space are
investigated. A convenient form of the metric is found and it is shown that in
each case the internal space is asymptotic to a cone over a product  of
spheres. A consistent truncation gives   gauged supergravities with
non-compact gauge groups. The BPS domain wall solutions of the non-compact
gauged supergravities are lifted to warped solutions in 10 or 11
dimensions.}
\endpage

\chapter{Non-Compact Gaugings and Higher Dimensional Solutions}

Gauged supergravities with compact gauge groups typically 
arise from dimensional reduction of 
higher dimensional supergravities with compact internal spaces.
In [\CW], it was shown that 
the supergravities  with non-compact gauge groups are associated with
higher dimensional supergravity solutions that
have a non-compact \lq internal' space.
In particular, 
de Sitter space solutions in $D=4,5$ arise in this way.
The no-go theorems of [\gib,\mn] imply that de Sitter space cannot arise
form a compactification of a higher-dimensional supergravity theory, and
the solutions of [\CW] get around this by having a non-compact internal
space. Our purpose here is to analyse the solutions of [\CW], and 
to generalise them to obtain the 11-dimensional origin of the solutions
of [\wall].

In $D=4$ the Cremmer-Julia $N=8$ supergravity theory [\CJ], with scalars
in the coset space $E_7/SU(8)$,  can be gauged by promoting a subgroup of
the rigid
$E_7$ symmetry to a local symmetry. 
The gauge group is necessarily 28-dimensional (as there are 28 vector
fields in the $N=8$  supermultiplet) and is always contained in the
$SL(8,\R)$  subgroup of
$E_7$ which is a symmetry of the ungauged action. The
$SO(8)$ gauging of [\dwn] has a maximally supersymmetric anti-de Sitter
vacuum. In [\nct,\nctt], gaugings with gauge group
$CSO(p,q,r)$ were obtained for all non-negative integers $p,q,r$ with
$p+q+r=8$, where
 $CSO(p,q,r)$ is  the group contraction of $SO(p+r,q)$
 preserving a symmetric metric with $p$ positive eigenvalues, $q$
negative ones and $r$ zero eigenvalues. Then  $CSO(p,q,0)=SO(p,q)$ and
$CSO(p,q,1)=ISO(p,q)$, and
$CSO(p,q,r)$  is not semi-simple if $r>0$. In [\frefo], it was argued that
these are the only possible gauge groups. Note that despite the
non-compact gauge groups, these are unitary theories. Of these
theories, the ones with gauge groups
$SO(4,4)
$ and
$SO(5,3)$ have de Sitter vacua arising at local maxima of the potentials 
[\nctt,\DS].  The $CSO(2,0,6)$ gauging has a Minkowski space solution
and the potential has flat directions [\nctt].
The structure and potentials of these models were analysed further in
[\HW]; no other critical points are known. In $D=5$, the gauged $N=8$ supergravities include
those with gauge groups
$SO(p,6-p)$ [\gunwar,\PPV] and of these the $SO(3,3)$ gauged theory has a
de Sitter vacuum.
In $D=7$,
 there 
are supergravities with gauge groups
$SO(p,5-p)$ [\PPVa].

The higher-dimensional origin of these theories was
found in [\CW]. Whereas the internal spaces for the compact gaugings are spheres, for the non-compact
gaugings, they are typically non-compact spaces of negative curvature.
Consider a solution of 
a $D$-dimensional gauged supergravity in which the only non-vanishing
fields are a  metric $ \bar g_{\mm\nn}(x) $ on a space-time
$M$ and certain scalar
fields $\phi(x)$. Then  these lift to solutions of
$d$-dimensional supergravity 
where $d=11$ for $D=4,7$, while for $D=5$ the solution lifts to a
solution of $d=10$ IIB supergravity.
In each of these cases, 
the metric is of the form
$$ ds^2= V^a \bar g_{\mm\nn}(x)dx^\mm
dx^\nn + V^b   g_{mn}(x,y) dy^m dy^n
\eqn\metric$$
where $  g_{mn}(x,y)$ is a metric on an \lq internal' space $N$ with
coordinates $y^m$ and dimension $d-D$,   $V(x,y)$ is a    warp factor and
$a,b$ are constants.
For solutions with constant scalars $\phi(x)$ at a critical point of the scalar
potential, the anti-symmetric tensor field strength is given in terms of the
volume forms on $M$ or $N$.
For $D=4$, the $d=11$ field strength $F_4$ is proportional to
the volume form on
$M$, for $D=5$ the IIB $F_5$ is a self-dual combination of the volume
forms on
$M$ and $N$, and for $D=7$, $*F_4$  is proportional to
the volume form on
$M$.
For solutions with varying scalars, the ansatz for the field strength is a little more complicated.

The compact  $SO(p)$ gaugings arise when the internal space $N$ is a
sphere
$S^{p-1}$. If all the scalars vanish, then $g_{mn}$ is the round metric
and the warp factor is 
$V=1$, but
non-vanishing scalars lead to a squashed metric on the sphere and 
a non-trivial warp factor [\ans,\consis].
 For the $CSO(p,q,r)$ gauging, the space $N$ is 
$\H ^{p,q,r}$ where $\H^{p,q,r}$ is a hypersurface of $\R^{p+q+r}$,
with a {\it positive definite metric } induced from the Euclidean metric
in
$\R^{p+q+r}$.
The hypersurface in $\R^{p+q+r}$ is that 
 in
which the real Cartesian coordinates $z^A$ of 
$\R^{p+q+r}$ satisfy
$$\eta _{AB}z^A z^B=R^2
,
\eqn\surf$$
where $\eta_{AB}$ is a metric with $p$ positive eigenvalues, $q$ negative
ones and $r$ zero eigenvalues. The scale  $R$ is determined by
the flux of an antisymmetric tensor gauge field strength. For example,
for $D=4,d=11$, the 4-form field strength of 11-dimensional supergravity
is proportional to $R$ times the volume form on $M$.
 Then   $\H ^{p,0,0}$ is a sphere $S^{p-1}$, 
  $\H ^{p,1,0}$ is the hyperboloid $H^p$, which is the coset
space
$SO(p,1)/SO(p)$,  and 
$\H ^{p,q,0}$ is a hyperbolic space (a non-symmetric space with negative
curvature)
 and $\H ^{p,q,r}=\H ^{p,q,0}\times \R^r$ is a generalised cylinder with
cross-section $\H ^{p,q}\equiv \H ^{p,q,0}$. For the cylinders, the flat
directions can be compactified to give
$\H ^{p,q,0}\times T^r$. 

It will be useful to consider metrics of the form
$$ \eta_{AB} =\pmatrix{
\II_{p \times p} & 0 &0 \cr
0 &  \xi \ll \II_{q \times q} &0\cr 0& 0& \zeta \II _{r\times r}
} ,\eqn\etis$$   
where we will usually take $\zeta =0$.\footnote*{The notation
for various quantities $V,\ll,R$  used here are related to that in
[\CW] as follows:
$V=\mm$, $\ll= c^2$, $R=r$.}
 Then the hypersurface \surf\ is a cylinder
$\R^r\times K$ (or $T^r\times K$)
with $p+q-1$ dimensional cross-section $K$.
If $\ll =1$ and $\xi=1$, then $K$ is a sphere $K=S^{p+q-1}$ with a \lq
round' metric, while if  $\ll =1$ and $\xi=-1$, then $K$ is the
hyperbolic surface
$\H^{p,q}$. If $\ll \ne 1$, then the resulting metric 
on $S^{p+q-1}$ or  $\H^{p,q}$ 
is \lq squashed'.
 The warp factor
(with $\zeta=0$) is
given by
$$V^2 =R^{-2}\left[ \sum _{i=1}^p (z^i)^2 +\xi^2 \ll^2 \sum _{i=p+1}^{p+q}
(z^i)^2\right]
.
\eqn\abc$$

 The simplest case is that in which all scalars $\phi$ are constant
and $(M,\bar g)$
 is an Einstein space with cosmological constant $\lll$.
Then the metric $\eta_{AB}$ is a constant matrix, $R$ is also a
constant and the warp factor  is a function of $y^m$ only, $V(x,y)=V(y)$.
The more general case in which some of the scalars are not constant
will be considered in   section 4, giving rise to a metric $g_{mn}(x,y)$ on
$N$ which varies with the coordinates $x$ on $M$.
 
Such solutions with $\lll<0$  arise for
the   compact gaugings in $D=4,5,7$, 
while solutions
with $\lll>0$ arise for
for the $D=4$ gaugings with gauge groups ${\cal G}=SO(4,4)$ and
$SO(5,3)$, and for  the
$D=5$ gauging with  ${\cal G}=SO(3,3)$.
There are also solutions with $\lll=0$, for example in the $D=4$ gauging
with ${\cal G}=SO(2,0,6)$, but these will not be discussed here.
For the other gaugings, there are no Einstein space solutions with
constant scalars, but there are BPS domain wall solutions with one or more
non-constant scalars [\wall], and these will be considered in  section 4.

The isometry group of the internal space is
the  unbroken gauge symmetry in the corresponding
gauged supergravity solution.
In a non-compact gauging, the gauge group is spontaneously broken to a
compact subgroup, and in each of the cases 
considered here, the isometry group of the internal space is compact.
This in particular implies that these spaces cannot be compactified by
identifying under the action of a discrete isometry group, although
new non-compact solutions can be obtained in some cases by such discrete
identifications.
The de Sitter solutions necessarily have no Killing spinors and so
completely break supersymmetry.
 The isometry group of
$\H^{p,q}$ is
$SO(p)\times SO(q)$, and  for the de Sitter solutions
with $(p,q)=(4,4),(5,3)$ or $(3,3)$, the $SO(p,q)$ gauge group
is broken to $SO(p)\times SO(q)$.
Similarly, for the inhomogeneously squashed $n$-sphere compactifications
of [\vanW] discussed below, the isometry group is $SO(n)$ and the 
corresponding gauged supergravity solutions have scalar
expectation values that break the $SO(n+1)$ 
gauge symmetry to $SO(n)$.

The solutions of the form \metric\ are as follows [\CW].
 In $D=4$, 
the solutions all have
$$a=2/3,
\qq b=-1/3, \qq \lll ={4\ll \over R^2}
.
\eqn\abc$$
The   de Sitter solutions with $\lll>0$ and $\xi=-1$
of the $SO(4,4)$ and $SO(5,3)$ gauged theories have 
$$p=4,\qq q=4, \qq \xi =-1 , \qq \ll =1
\eqn\abc$$ 
and
$$p=5,\qq q=3,\qq \xi =-1 , \qq \ll =3
,
\eqn\abc$$ 
respectively.
There are also two anti-de Sitter solutions of this form with $\lll<0$
arising for the compact $SO(8)$ gauging with $\xi=1$; these are the round
seven sphere solution $AdS_4\times S^7$ and 
 the inhomogeneously squashed 7-sphere compactification of [\vanW], which
is the solution with
$$p=7,\qq q=1, \qq \xi =1 , \qq \ll =-5
.
\eqn\abc$$

In $D=5$, 
the solutions all have
$$ a=1/2, \qq b=-1/2, \qq \lll ={2\ll \over R^2}
.
\eqn\abc$$
The de Sitter solution of the $SO(3,3)$ gauged theory in $D=5$
arises from the   solution of IIB  supergravity with $D=5$ and
$$p=3,\qq q=3,\qq \ll =1
,
\eqn\abc$$ and with the self-dual 5-form field strength given in terms of
the volume forms on $dS_5$ and the internal space.
There are also two $\lll<0$ solutions for the compact $SO(6)$ gauging, the
round
$S^5$ solution and the 
inhomogeneously squashed 5-sphere compactification of [\vanW], which
is the solution with
$$p=5,\qq q=1, \qq \xi =1 , \qq \ll =-3
\eqn\abc$$

In $D=7$, 
the solutions all have
$$ a=1/3, \qq b=-2/3, \qq \lll ={\ll \over R^2}
.
\eqn\abc$$
There are no $\lll>0$ solutions, but there   are two $\lll<0$ solutions
for the compact gauging,
the round 4-sphere solution and the
inhomogeneously squashed 4-sphere compactification of [\vanW], which
is the solution with
$$p=4,\qq q=1, \qq \xi =1 , \qq \ll =-2
.
\eqn\abc$$

\chapter{The Metric on the Internal Space }

To obtain a more useful form of the metric on
the hyperboloid $\H ^{p,q}$, 
introduce first coordinates $\ss, \th_i,\ti \ss,\ti \th_i$ on $\R^{p+q}$,
where $\th_i$ are coordinates on $S^{p-1}$, $\ti \th_i$ are coordinates on
$S^{q-1}$ and $\ss,\ti \ss$ are radial coordinates on $\R^p$ and $\R^q$,
respectively.
The Euclidean metric on $\R^{p+q}$
is then
$$ ds^2= d\ss^2 + \ss^2 d\www_{p-1}^2+
 d\ti \ss^2 +\ti \ss^2 d\www_{q-1}^2
\eqn\euc$$
where $d\www_n^2$ is the round metric on $S^{n}$.
Next, we introduce coordinates
$r,\rho$ with
$$ \ti \ss=\rho r, \qq  \ss=\rho (1+\ll  r^2)^{1/2}
,
\eqn\abc$$
so that
$$ \ss^2-\ll \ti \ss^2=\rho^2
.
\eqn\abc$$
Then the hypersurface $\H ^{p,q}$ is defined by $\rho =R$ and
$r,\th_i, \ti \th _i$ are  intrinsic coordinates
on the space, and the metric on $\H ^{p,q}$ induced by \euc\
is given by transforming   to coordinates $\rho,r,\th_i, \ti \th _i$
and setting $\rho=R$. This gives
$$
ds^2 =R^2\left( {1+ (\ll  +\ll^2)r^2 \over 1+ \ll  r^2}
dr^2+
 (1+ \ll  r^2)
  d\www_{p-1}^2+
r^2 d\www_{q-1}^2
\right)
.
\eqn\met$$
The warp factor is
$$V=R^{-2}\left(1 + (1+\ll^2)r^2\right)
\eqn\vis$$
so that the metric on $N$ is \met\ times $V^b$.

At large $r$, the metric takes the asymptotic form
$$
ds^2=(1+\ll ) R^2  \left( dr^2
+\aa r^2
  d\www_{p-1}^2
+\bb r^2 d\www_{q-1}^2
\right)
\eqn\abc$$
where
$$
\aa= {\ll \over 1+\ll },
\qq
\bb = {1\over 1+\ll }
,
\eqn\abc$$
so that asymptotically this is a cone over $S^{p-1}\times S^{q-1}$, with
a product metric that is not Einstein for $q>1$. Asymptotically, the warp
factor becomes
$$V=R^{-2}  (1+\ll^2)r^2 
.
\eqn\abc$$
The distance from an interior point  to the
boundary  at
$r
\to
\infty$ 
is infinite for the metric \met,
and   remains infinite for the metric warped by $V^b$
if $$b>-1
\eqn\abc$$
For all the solutions considered here, $b>-1$ and the internal space has
infinite volume.

For $\ll =1$, the metric \met\ is
$$
ds^2 =R^2\left( {1+ 2r^2 \over 1+    r^2}
dr^2+
 (1+    r^2)
  d\www_{p-1}^2+
r^2 d\www_{q-1}^2
\right)
\eqn\met$$
and the warp factor is
$$V=R^{-2}\left(1 + 2r^2\right)
.
\eqn\abc$$
This is the geometry of the $D=4$, $SO(4,4)$ gauging with $p=q=4$ and the
$D=5$, $SO(3,3)$ gauging with $p=q=3$, while that for
the $D=4$, $SO(5,3)$ gauging is a squashed version of this with
$\ll=3$.

It is interesting to note that the radius of the $S^{p-1}$ never
vanishes.
This means that one may identify points under the action
of a discrete subgroup $\Gamma \subset SO(p)$ and as long as $\Gamma$
acts freely on $S^{p-1}$, i.e. has no fixed points, the quotient will
be smooth. In general this will break the $SO(p)$ factor
of the  gauge group down to
that continuous  subgroup of $SO(p)$ which commutes with $\Gamma$.
In some cases this may break $SO(p) \times SO(q)$ to $SO(q)$.
In this way we could eliminate some Klauza-Klein
gauge states. However since the radius of the $S^{q-1}$ vanishes
at the origin $r=0$ we cannot use this mechanism to break the $SO(p)$
factor.

\chapter{A Lower-Dimensional Interpretation?}

With a non-compact internal space, an important issue is whether the theory
is instrinsically higher dimensional, or whether the physics
  can be consistently interpreted as $D$ dimensional, perhaps through a
  brane-world scenario [\randsum] or a
  Kaluza-Klein-like dimensional reduction with non-compact space, as in 
  [\gel,\NicWet].
 In particular, because the internal space is
non-compact there appears to be a difficulty in obtaining conventional
4-dimensional gravity for sources which are located at fixed values of the
internal coordinate $y$. For instance one could imagine sources localized
near the centre $r=0$. The conventional criterion for  getting
gravity in this way is obtained by computing from the action the
effective  $D$-dimensional Newton's constant $G_D$,
giving $G_D=G_d/ {\cal V}$, where $G_d$ is the $d$-dimensional Newton's constant
and ${\cal V}$
   is  the volume of
  the internal manifold $N$ with respect to the metric $V^bg_{mn}$. 
  For instance in
the case $p=q=4, b=-{ 1\over 3}, \lambda =1$ the integral is
$$
{\cal V}= \int d ^7 y \sqrt{g} V^{ 7 b \over 2} = \int d^7 y \sqrt{g} V^{-{
7 \over 6}}.
\eqn\wavasd
$$
Using the formulae for the metric and warp factor this integral is
easily seen to diverge,  and it also does so for the other cases [\CW], with the
result that the effective gravitational coupling
$G_D=0$ in each case. This
indicates that the Randall-Sundrum mechanism [\randsum] should not apply to this case, and
  gravity is not localised in the large extra dimensions of the internal
space; indeed, we have checked that there are no graviton modes that are normalizable with respect to the
internal space.

An alternative view point might be to consider sources which are
in some sense delocalized over the internal space. Modes that are constant
over the internal space would be
 described by the consistent truncation of the
higher dimensional
 supergravity equations to the lower dimensional gauged
supergravity equations. This would be to discard all solutions
with non-trivial dependence on the internal  
coordinates. While mathematically consistent it is physically more
appealing to consider non-trivial dependence and to seek boundary
conditions on the internal manifold which would allow a lower
dimensional interpretation. Note that despite the vanishing of the effective
Newton's constant $G_D$ and the difficulty of reducing the action, dimensionally
reducing the field equations is straightforward and involves no divergent
integrals [\DS].

To see what is involved in more detail, we consider those
graviton modes in $d$ dimensions which take the form  $ \psi  (y) 
h_{\mu \nu}(x)$ where $h_{\mu \nu}$ are   $D$-dimensional graviton 
modes and $\psi$ satisfies the scalar wave equation with respect 
to the higher dimensional metric, with the warp factor taken into 
account. 
 Assuming only dependence on $r$ we find for the $D=4$, $SO(4,4)$ case  that 
$$ 
 { (1+2r^2 ) ^{ 2 \over 3} \over (1+r^2 ) r^3 } \; { d \over dr} 
 \Bigl ( { (1+r^2 ) r^3 \over ( 1+ r^2 ) ^{ 4 \over 3}} \, { d \psi \over dr }
\Bigr ) = -\Lambda \psi,  \eqn\wave
$$ 
where $\Lambda$ is a separation constant. The equation is 
self-adjoint with respect to the inner product 
$$ 
\int |\psi |^2 dr r^3 (1+r^2) ( 1+ 2 r^2) ^{-2 \over 3}. 
\eqn\wafssf
$$ 
Note that the norm for constant $\psi$, (with $\Lambda =0$) is 
proportional to the volume integral giving ${\cal V}$. 

The problem now is whether one can find boundary conditions to fix
the values of the separation constant $\Lambda$ to give a discrete
spectrum. By looking at the behaviour of the solutions near
infinity this looks doubtful. If $\Lambda$ is negative  we get
solutions which either blow up or decay exponentially. If we pick
the latter,  then  multiplying the equation \wave\ by $\psi$ and integrating
by parts, dropping a surface term which should vanish with the exponential
fall-off, then  we obtain a contradiction, as a manifestly positive integral
is set equal to one that is manifestly negative for any solutions which fall
off sufficiently fast at infinity to allow the dropping of the surface term.
We conclude that $\lll$ cannot be negative for such boundary conditions.  If
$\Lambda $ is positive the solutions oscillate near infinity and and are not 
normalizable; in this regime, we expect a continuous spectrum for $\lll$. It
seems therefore that one obtains similar difficulties to those encountered in
[\Gibbons, \Gibbwilt]. This is a pity since, as remarked in [\Gibb] a picture
of this sort could  explain in natural way the
 dimensionality of our
spacetime.

The non-compact gauged supergravities can then be obtained from a consistent
truncation of 
the higher dimensional theory to those
  modes that are constant on the internal space. 
  Without the truncation, the $D$-dimensional spectrum appears to consist of the 
  gauged supergravity together with a continuous spectrum of massive states
  arising from
  non-normalizable modes on $N$, and
  there does not
seem to be any natural way to impose  boundary conditions or otherwise restrict
the theory in such a way
that a $D$-dimensional spectrum with a mass gap emerges. In this sense, the
theory is more naturally viewed from the $d$-dimensional viewpoint.

\chapter{Domain Wall Solutions from Higher Dimensions}

Any solution of a gauged supergravity should have
a lifting to $d=10$ or $d=11$ of the kind discussed above, although the
ansatz will be more complicated for solutions with more non-vanishing
  fields.
Most of the gauged supergravity theories do not
have de Sitter or anti-de Sitter solutions, but most have
half-supersymmetric
domain wall solutions with $D-1$ dimensional Poincar\' e symmetry
[\wall,\LPS-\Ahn]. In these, the metric has the form
$$ds^2= e^{2 A(u)} ds^2\left(\bbbb {E}^{(1,D-2)}\right) +e^{2B(u)} du^2
\eqn\wallmet$$
and the functions $A,B$ and the scalars $\phi(u)$ are functions of the
transverse coordinate $u$ only. We will now consider the lifting of these
to $d=10$ or $d=11$ solutions.

Consider the subset $SL(n,\R)/SO(n)$ of the scalar coset space, where
$n=8$ for $D=4$, $n=6$ for $D=5$ and $n=5$ for $D=7$.
(For $D=7$, this is the whole scalar coset space, while for $d=4,5$ it is
a subspace.)
A  configuration of scalar fields from this subspace is then represented
by an $SL(n,\R)$ matrix $S_A{}^B(x)$.
In the compact $SO(n)$ gauging, such a scalar configuration leads to a
squashing of the $n-1$ sphere to the surface \surf\ in $\R^n$
with
$$ \eta _{AB} = S_A{}^CS_B{}^D\dd_{CD}
\eqn\abc$$
and a warp factor
$$V=\eta ^{AC} \eta ^{AD}z^Cz^D
.
\eqn\abc$$
The ansatz for the 4-form  [\ans,\consis] or 5-form field strength $F$ is also given in terms of
$S(x)$.

This can then be analytically continued to the non-compact gaugings [\CW].
For any $p,q$ with $p+q=n$ there is an
$SO(p)\times SO(q)$ invariant direction in the scalar coset space
represented by matrices
of the form
$$S_{p,q}(t) 
=\pmatrix{
e^{t/2} \II_{p \times p} & 0   \cr
0 & e^{-\bb t/2}   \II_{q \times q} }
\eqn\abc$$
where
$\bb=p/q$ and $t$ is a parameter.
For any finite $t$, acting with the $SL(n,\R)$ transformation
$S_{p,q}(t)$ and rescaling the coupling constant
$$g \to ge^{-t}
\eqn\abc$$
defines a one-parameter family
of theories, all equivalent  to the compact gauging through a field
redefinition.
Let 
$$\xi = \exp [-(1+p/q)t]
.
\eqn\abc$$
For the $D=4$ theory,
continuing to $t=\infty, \xi=0$ or to $t=-iq\pi /(p+q), \xi=-1$
define new theories, the $CSO(p,0,q)$ gauging  at $\xi=0$ and the 
$SO(p,q)$ gauging at $\xi=-1$. For $D=5,7$,  
non-compact semi-simple gaugings   arise in a similar  way, and the
$SO(p,q)$ gauging is associated with the continuation to
$\xi=-1$.

It was seen in [\CW] that this
continuation could be applied to the $d$-dimensional theory, with the
$SL(n,\R)$ acting naturally on the space $\R^n$ in which the sphere
$S^{n-1}$ or hypersurface $\H^{p,q,r}$ (with $p+q+r=n-1$) is embedded. 
The ansatz for the 4-form or 5-form field strength is also transformed by the
$SL(n,\R)$ transformation, but we will only display the transformed metric here.
Note that the continuation takes solutions of the complexified field
equations of the compact gauging to solutions of the complexified field
equations of the non-compact gaugings, but does not in general take real
solutions to real solutions. 

For scalar configurations $S(x)$ such that $S(x)$  commutes with
$S_{p,q}(t)$, it is straightforward to calculate the effect of such
continuations on the potential [\nctt] or the $d$-dimensional geometry
[\CW]. (For more general configurations, see [\HW].)
The hypersurface   becomes \surf\ with
$$ \eta _{AB} = \hat S_A{}^C \hat S_B{}^D\dd_{CD}
\eqn\etist$$
where
$$ \hat S= e^{-t/2}S_{p,q}(t) S(x)
.
\eqn\abc$$
Consider the case in which $S(x)$ is $SO(p)\times SO(q)$ invariant
$$S (x)=\pmatrix{
e^{\phi/2} \II_{p \times p} & 0   \cr
0 & e^{-\bb \phi/2}   \II_{q \times q} }
.
\eqn\abc$$
where
$\phi(x)$ is a scalar field.
Configurations with constant $\phi$
 were considered explicitly in [\CW], but the generalisation to
non-constant fields is straightforward.
The resulting geometry of the internal space $N$ now depends on the
position in $M$, so that the internal  metric $g_{mn}$ 
depends on both $x$ and $y$.
The hypersurface is now \surf\ with
$$\eta _{AB}=e^\phi  \pmatrix{
  \II_{p \times p} 
& 0   \cr
0 &   \xi e^{-(1+\bb)  \phi}\II_{q \times q} 
}
\eqn\etvar$$
The surface can then be written as
$$e^{-\phi} R^2 =   \sum _{i=1}^p (z^i)^2 +\xi  e^{-(1+\bb) \phi}  \sum
_{i=p+1}^{p+q} (z^i)^2
,
\eqn\abc$$ 
which is
of the form \surf\ with
$$ \eta_{ab} =\pmatrix{
\II_{p \times p} & 0   \cr
0 &  \xi \ll \II_{q \times q}  
}\eqn\etiss$$ 
 and where now the moduli
$R, c$ are replaced with functions of $x$:
$$ R(x)= e^{-\phi/2} R, \qq \ll (x)= e^{-(1+\bb) \phi}
\eqn\ris$$
for  some constant $R$.
The metric on this hypersurface   is now \met\ with these $x$-dependent 
$R, c$, so that the radius $R$ and the squashing parameter $\ll $
both vary with $x$.
The warp factor is \vis, but with $R,\ll$ replaced by $R(x),\ll(x)$ given
by \ris.

For $D=4$, the $CSO(p,q,r)$ gaugings can be obtained from the $SO(p,q+r)$
gaugings by a further field continuation in the $SO(p+q)\times SO(r)$
invariant direction, using $S'(s)=S_{p+q,r}(s)$.
Then the hypersurface becomes \surf,\etist\  with
$$ \hat S= e^{-(t+s)/2}S_{p,q+r}(t)S_{p+q,r}(s) S(x)
,
\eqn\abc$$
giving a 2-parameter set of theories depending on 
$$\xi = \exp \left[-\left (1+{p\over q+r} \right) t \right ],
\qq
\zeta
= \exp \left[-\left (1+{p+q\over r} \right) s \right ]
.
\eqn\abc$$
The $CSO(p,q,r)$ gauging arises at
$\xi=-1,\zeta=0$, where
$$\eta _{AB}=e^{\phi+\chi} 
\pmatrix{
  \II_{p \times p} 
& 0  & 0  \cr
0 &  - e^{-(1+\bb)  \phi}\II_{q \times q} 
&0\cr 0& 0& {  0} _{r\times r}
}
,
\eqn\abc$$
so that the metric is the metric \met\ on $\H^{p,q,r}$ with
$$ R(x)= e^{-(\phi +\chi)} R, \qq \ll (x)= e^{-(1+\bb) \phi}
\eqn\riss$$
with warp factor given in terms of $R(x),\ll(x)$ by \vis.

The domain wall solutions of [\wall], with metric \wallmet\ for particular
$A(u),B(u)$ and scalar $\phi(u)$,
 are then lifted to
$d=10$ or $d=11$ dimensional solutions with metric \metric,\met\ and
$R(u),\ll(u)$ given by \ris\ or \riss, and warp factor \vis.
It is straightforward to obtain the relevant antisymetric gauge field
for these solutions by transforming the standard ansatz.

\refout

\bye